\documentclass{llncs}

\usepackage{graphicx}
\usepackage{amsmath}

\newcommand{\ea}[1]{#1 \emph{et al.}}

\begin{document}
 
\title{Exploring chance in NCAA basketball}
\author{Albrecht Zimmermann\\\email{albrecht.zimmermann@insa-lyon.fr}}
\institute{INSA Lyon}

\maketitle

\begin{abstract}
 There seems to be an upper limit to predicting the outcome of matches in (semi-)professional sports. Recent work has proposed that this is due to chance and attempts have been made to simulate the distribution of win percentages to identify the most likely proportion of matches decided by chance. We argue that the approach that has been chosen so far makes some simplifying assumptions that cause its result to be of limited practical value. Instead, we propose to use clustering of statistical team profiles and observed scheduling information to derive limits on the predictive accuracy for particular seasons, which can be used to assess the performance of predictive models on those seasons. We show that the resulting simulated distributions are much closer to the observed distributions and give higher assessments of chance and tighter limits on predictive accuracy.
\end{abstract}

\section{Introduction}

In our last work on the topic of NCAA basketball \cite{zimmermann2013predicting}, we speculated about the existence of a ``glass ceiling'' in (semi-)professional sports match outcome prediction, noting that season-long accuracies in the mid-seventies seemed to be the best that could be achieved for college basketball, with similar results for other sports. One possible explanation for this phenomenon is that we are lacking the attributes to properly describe sports teams, having difficulties to capture player experience or synergies, for instance. While we still intend to explore this direction in future work,\footnote{Others in the sports analytics community are hard at work doing just that, especially for ``under-described sports such as European soccer or NFL football.} we consider a different question in this paper: \emph{the influence of chance on match outcomes}.

Even if we were able to accurately describe sports teams in terms of their performance statistics, the fact remains that athletes are humans, who might make mistakes and/or have a particularly good/bad day, that matches are refereed by humans, see before, that injuries might happen during the match, that the interaction of balls with obstacles off which they ricochet quickly becomes too complex to even model etc. Each of these can affect the match outcome to varying degrees and especially if we have only static information from before the match available, it will be impossible to take them into account during prediction. 

While this may be annoying from the perspective of a researcher in sports analytics, from the perspective of sports leagues and betting operators, this is a feature, not a bug. Matches of which the outcome is effectively known beforehand do not create a lot of excitement among fans, nor will they motivate bettors to take risks.

Intuitively, we would expect that chance has a stronger effect on the outcome of a match if the two opponents are roughly of the same quality, and if scoring is relatively rare: since a single goal can decide a soccer match, one (un)lucky bounce is all it needs for a weaker team to beat a stronger one. In a fast-paced basketball game, in which the total number of points can number in the two hundreds, a single basket might be the deciding event between two evenly matched teams but probably not if the skill difference is large.

For match outcome predictions, a potential question is then: ``\emph{How strong is the impact of chance for a particular league?}'', in particular since quantifying the impact of chance also allows to identify the ``glass ceiling'' for predictions. The topic has been explored for the NFL in \cite{burke07LuckNFL01}, which reports
\begin{quote}
 The actual observed distribution of win-loss records in the NFL is indistinguishable from a league in which 52.5\% of the games are decided at random and not by the comparative strength of each opponent.
\end{quote}
Using the same methodology, \ea{Weissbock} \cite{DBLP:conf/ai/WeissbockI14} derive that 76\% of matches in the NHL are decided by chance. As we will argue in the following section, however, the approach used in those works is not applicable to NCAA basketball.

\section{Identifying the impact of chance by Monte Carlo simulations}

The general idea used by Burke and Weissbock\footnote{For details for Weissbock's work, we direct the reader to \cite{weissbock13mlForNHL02}.} is the following:
\begin{enumerate}
 \item A chance value $c \in [0,1]$ is chosen.
 \item Each out of a set of virtual teams is randomly assigned a strength rating.
 \item For each match-up, a value $v \in [0,1]$ is randomly drawn from a uniform distribution.
 \begin{itemize}
  \item If $v \geq c$, the stronger team wins.
  \item Otherwise, the winner is decided by throwing an unweighted coin.
 \end{itemize}
 \item The simulation is re-iterated a large number of times (e.g. $10,000$) to smooth results.
\end{enumerate}

\begin{figure}[ht]
\centering
 \includegraphics[angle=270,width=0.7\linewidth]{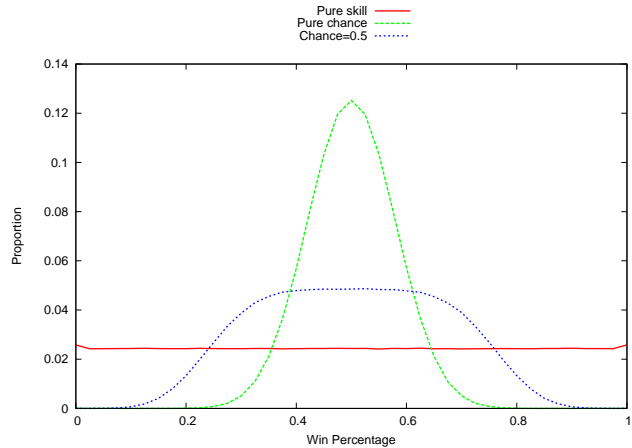}
 \caption{MC simulated win percentage distributions for different amounts of chance\label{pure-curves}}
\end{figure}

Figure \ref{pure-curves} shows the distribution of win percentages for $340$ teams, $40$ matches per team (roughly the settings of an NCAA basketball season including playoffs), and $10,000$ iterations for $c = 0.0$ (pure skill), $c = 1.0$ (pure chance), and $c=0.5$.

By using a goodness of fit test -- $\chi^2$ in the case of Burke's work, \emph{F-Test} in the case of Weissbock's -- the $c$-value is identified for which the simulated distribution fits the empirically observed one best, leading to the values reproduced in the introduction. The identified $c$-value can then be used to calculate the upper limit on predictive accuracy in the sport: since in $1-c$ cases the stronger team wins, and a predictor that predicts the stronger team to win can be expected to be correct in half the remaining cases in the long run, the upper limit lies at:
$$(1-c) + c/2\text{,}$$
leading in the case of
\begin{itemize}
 \item the NFL to: $0.475 + 0.2625 = 0.7375$, and
 \item the NHL to: $0.24 + 0.36 = 0.62$
\end{itemize}
Any predictive accuracy that lies above those limits is due to the statistical quirks of the observed season: theoretically it is possible that chance always favors the stronger team, in which case predictive accuracy would actually be $1.0$. As we will argue in the following section, however, NCAA seasons (and not only they) are likely to be quirky indeed.

\section{Limitations of the MC simulation for NCAA basketball}
\label{limitations}

A remarkable feature of Figure \ref{pure-curves} is the symmetry and smoothness of the resulting curves. This is an artifact of the distribution assumed to model the theoretical distribution of win percentages -- the Binomial distribution -- together with the large number of iterations. This can be best illustrated in the ``pure skill'' setting: even if the stronger team were always guaranteed to win a match, real-world sports schedules do not guarantee that any team actually plays against representative mix of teams both weaker and stronger than itself. A reasonably strong team could still lose every single match, and a weak one could win at a reasonable clip. One league where this is almost unavoidable is the NFL, which consists of 32 teams, each of which plays 16 regular season matches (plus at most 4 post-season matches), and ranking ``easiest'' and ``hardest'' schedules in the NFL is an every-season exercise. Burke himself worked with an empirical distribution that showed two peaks, one before $0.5$ win percentage, one after. He argued that this is due to the small sample size (five seasons).

\begin{figure}[ht]
\centering
 \includegraphics[angle=270,width=0.7\linewidth]{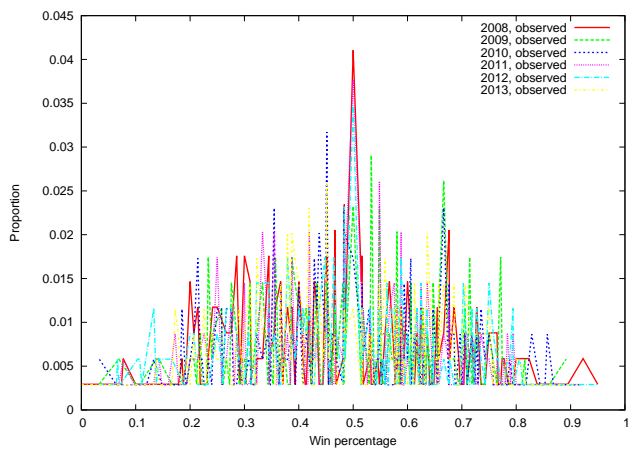}
 \caption{Observed distribution of win percentages in the NCAA, 2008--2013\label{observed-ncaa-curves}}
\end{figure}
The situation is even more pronounced in NCAA basketball, where 340+ Division I teams play at most 40 matches each. Figure \ref{observed-ncaa-curves} shows the empirical distribution for win percentages in NCAA basketball for six season (2008--2013).\footnote{The choice of seasons is purely due to availability of data at the time of writing and we intend to extend our analysis in the future.} While there is a pronounced peak for a win percentage of $0.5$ for 2008 and 2012, the situation is different for 2009, 2010, 2011, and 2013. Even for the former two seasons, the rest of the distribution does not have the shape of a Binomial distribution. Instead it seems to be that of a \emph{mix} of distributions -- e.g. ``pure skill'' for match-ups with large strength disparities overlaid over ``pure chance'' for approximately evenly matched teams. 

NCAA scheduling is subject to conference memberships and teams will try to pad out their schedules with relatively easy wins, violating the implicit assumptions made for the sake of MC simulations.
This also means that the ``statistical quirks'' mentioned above are often the norm for any given season, not the exception. Thought to its logical conclusion, the results that can be derived from the Monte Carlo simulation described above are purely theoretical: if one could observe an {\bf effectively unlimited} number of seasons, during which schedules are {\bf not systematically imbalanced}, the overall attainable predictive accuracy were bound by the limit than can be derived by the simulation. For a given season, however, and the question how well a learned model performed w.r.t. the specificities of that season, this limit might be too high (or too low).

\begin{figure}
\centering
 \includegraphics[angle=270,width=0.7\linewidth]{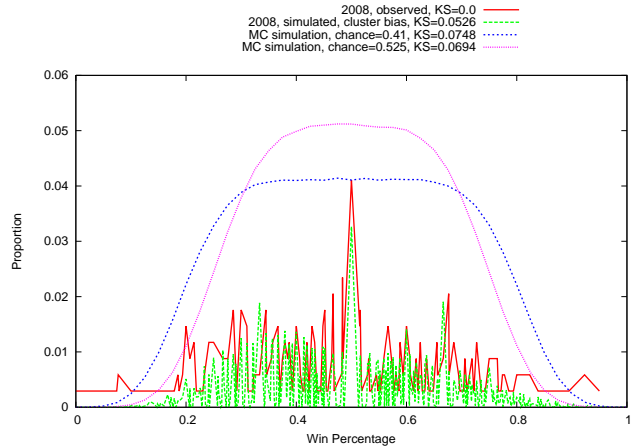}
 \caption{Distribution of win percentages 2008\label{2008}}
\end{figure}
As an illustration, consider Figure \ref{2008}.\footnote{Other seasons show similar behavior, so we treat 2008 as a representative example.} The MC simulation that matches the observed proportion of teams having a win percentage of $0.5$ is derived by setting $c=0.42$, implying that a predictive accuracy of $0.79$ should be possible. The MC simulation that fits the observed distribution best, according to the Kolmogorov-Smirnov (KS) test (overestimates the proportion of teams having a win percentage of $0.5$ along the way), is derived from $c=0.525$ (same as Burke's NFL analysis), setting the predictive limit to $0.7375$. Both curves have visually nothing in common with the observed distribution, yet the null hypothesis -- that both samples derive from the same distribution -- is not rejected at the 0.001 level by the KS test for sample comparison. This hints at the weakness of using such tests to establish similarity: CDFs and standard deviations might simply not provide enough information to decide whether a distribution is appropriate.

\section{Deriving limits for specific seasons}

The ideal case derived from the MC simulation does not help us very much in assessing how close a predictive model comes to the best possible prediction. Instead of trying to answer the theoretical question: \emph{What is the expected limit to predictive accuracy for a given league?}, 

\noindent we therefore want to answer the practical question: \emph{Given a specific season, what was the highest possible predictive accuracy?}. 

To this end, we still need to find a way of estimating the impact of chance on match outcomes, while \emph{taking the specificities of scheduling into account}. The problem with estimating the impact of chance stays the same, however: for any given match, we need to know the relative strength of the two teams but if we knew that, we would have no need to learn a predictive model in the first place. If one team has a lower adjusted \underline{offense} efficiency than the other (i.e. scoring less), for example, but also a lower adjusted \underline{defensive} efficiency (i.e. giving up fewer points), should it be considered weaker, stronger, or of the same strength?

Learning a model for relative strength and using it to assess chance would therefore feed the models potential errors back into that estimate. What we \emph{can} attempt to identify, however, is which teams are \emph{similar}.

\subsection{Clustering team profiles and deriving match-up settings}

\begin{table}[ht]
\centering
\begin{scriptsize}
 \begin{tabular}{c|c||c|c}
  \multicolumn{2}{c||}{Offensive stats}&\multicolumn{2}{c}{Defensive stats}\\\hline
  AdjOEff & Points per 100 possessions scored, & AdjDEff & Points per 100 possessions allowed,\\
  & adjusted for opponent's strength & & adjusted for opponent's strength\\\hline
  OeFG\% & Effective field goal percentage& DeFG\% & eFG\% allowed\\\hline
  OTOR & Turnover rate& DTOR& TOR forced\\\hline
  OORR& Offensive rebound rate&DORR& ORR allowed\\\hline
  OFTR&Free throw rate&DFTR& FTR allowed\\
 \end{tabular}
 \end{scriptsize}
\caption{Team statistics\label{profile-stats}}
\end{table}

We describe each team in terms of their adjusted efficiencies, and their Four Factors, adopting Ken Pomeroy's representation \cite{kenpom}. Each statistic is present both in its offensive form -- how well the team performed, and in its defensive form -- how well it allowed its opponents to perform (Table \ref{profile-stats}). We use the averaged end-of-season statistics, leaving us with approximately 340 data points per season. Clustering daily team profiles, to identify finer-grained relationships, and teams' development over the course of the season, is left as future work. As a clustering algorithm, we used the WEKA \cite{weka} implementation of the EM algorithm with default parameters. This involves EM selecting the appropriate number of clusters by internal cross validation, with the second row of Table \ref{number-clusters} showing how many clusters have been found per season.

\begin{table}[ht]
\centering
 \begin{tabular}{c|c|c|c|c|c|c}
  Season & 2008 & 2009 & 2010 & 2011 & 2012 & 2013\\\hline
  Number of Clusters & 5 & 4 & 6 & 7 & 4 & 3\\
  Cluster IDs in Tournament & 1,5 & 4 & 2,6 & 1,2,5 & 3,4 & 2\\
 \end{tabular}
\caption{Number of clusters per season and clusters represented in the NCAA tournament\label{number-clusters}}
\end{table}

As can be seen, depending on the season, the EM algorithm does not separate the 340 teams into many different statistical profiles. Additionally, as the third row shows, only certain clusters, representing relatively strong teams, make it into the NCAA tournament, with the chance to eventually play for the national championship (and one cluster dominates, like Cluster 5 in 2008). These are strong indications that the clustering algorithm does indeed discover similarities among teams that allow us to abstract ``relative strength''. Using the clustering results, we can re-encode a season's matches in terms of the clusters to which the playing teams belong, capturing the specificities of the season's schedule.

\begin{table}[ht]
\centering
 \begin{tabular}{c|c|c|c|c|c||c}
  & Cluster 1 & Cluster 2 & Cluster 3 & Cluster 4 & Cluster 5 & Weaker opponent\\\hline
  Cluster 1 & 76/114 & 161/203 & 52/53 & 168/176 & 65/141 & 381/687 (0.5545)\\
  Cluster 2&100/176&298/458&176/205&429/491&91/216&705/1546 (0.4560)\\
  Cluster 3&7/32&55/170&47/77&119/194&4/40&119/513 (0.2320)\\
  Cluster 4&22/79&161/379&117/185&463/769&28/145&117/1557 (0.0751)\\
  Cluster 5 & 117/154 & 232/280 & 78/83 & 232/247 & 121/198 & 659/962 (.6850)\\
 \end{tabular}
\caption{Wins and total matches for different cluster pairings, 2008\label{schedule}}
\end{table}

Table \ref{schedule} summarizes the re-encoded schedule for 2008. The re-encoding allows us to flesh out the intuition mentioned in the introduction some more: teams from the same cluster can be expected to have approximately the same strength, increasing the impact of chance on the outcome. Since we want to take all non-chance effects into account, we encode pairings in terms of which teams has home-court. The left margin indicates which team has home court in the pairing: this means, for instance, that while teams from Cluster 1 beat teams from Cluster 2 almost 80\% of the time when they have home court advantage, teams from Cluster 2 prevail in almost 57\% of the time if home court advantage is theirs. The effect of home court advantage is particularly pronounced on the diagonal, where unconditional winning percentages by definition should be at approximately 50\%. Instead, home court advantage pushes them always above 60\%. One can also see that the majority of cases teams were matched up with a team stronger than (or as strong as) themselves. Table \ref{schedule} is the empirical instantiation of our remark in Section \ref{limitations}: instead of a single distribution, 2008 seems to have been a weighted mixture of 25 distributions.\footnote{Although some might be similar enough to be merged.} None of these specificities can be captured by the unbiased MC simulation.

\subsection{Estimating chance}

The re-encoded schedule includes all the information we need to assess the effects of chance. The win percentage for a particular cluster pairing indicates which of the two clusters should be considered the stronger one in those circumstances, and from those matches that are lost by the stronger team, we can calculate the chance involved.

Consider, for instance, the pairing \emph{Cluster 5 -- Cluster 2}. When playing at home, teams from Cluster 5 win this match-up in 82.85\% of the cases! This is the practical limit to predictive accuracy in this setting for a model that always predicts the stronger team to win, and in the same way we used $c$ to calculate that limit above, we can now inverse the process: $c = 2*(1-0.8285) = 0.343$. When teams from Cluster 5 welcomed teams from Cluster 2 on their home court in 2008, the overall outcome is indistinguishable from 34.3\% of matches having been decided by chance.

The impact of chance for each cluster pairing, and the number of matches that have been played in particular settings, finally, allows us to calculate the effect of chance on the entire season, and using this result, the upper limit for predictive accuracy that could have been reached for a particular season.

\begin{table}
\centering
 \begin{tabular}{|c|c|c|c|c|c|c|}\hline
  Season & 2008 & 2009 & 2010 & 2011 & 2012 & 2013\\\hline
  \multicolumn{7}{|c|}{Unconstrained EM}\\\hline
  KS & 0.0526 & 0.0307 & 0.0506 & 0.0327 & 0.0539 & 0.0429\\
  Chance & 0.5736 & 0.5341 & 0.5066 & 0.5343 & 0.5486 & 0.5322 \\
  Limit for predictive accuracy & 0.7132 & 0.7329 & 0.7467 & 0.7329 & 0.7257 & 0.7339\\\hline
  \multicolumn{7}{|c|}{Optimized EM (Section \ref{optimization})}\\\hline
  KS & 0.0236 & 0.0307 & 0.0396 & 0.0327 & 0.0315 & 0.0410\\
  Chance & 0.4779 & 0.5341 & 0.4704 & 0.5343 & 0.4853 & 0.5311\\
  Limit & 0.7610 & 0.7329 & 0.7648 & 0.7329& 0.7573 & 0.7345\\\hline\hline
  KenPom prediction & 0.7105 & 0.7112 & 0.7244 & 0.7148 & 0.7307 & 0.7035\\\hline
 \end{tabular}
\caption{Effects of chance on different seasons' matches and limit on predictive accuracy (for team encoding shown in Table \ref{profile-stats})\label{limits}}
\end{table}

The upper part of Table \ref{limits} shows the resulting effects of chance and the limits regarding predictive accuracy for the six seasons under consideration. Notably, the last row shows the predictive accuracy when using the method described on \cite{kenpom}: the Log5-method, with Pythagorean expectation to derive each team's win probability, and the adjusted efficiencies of the home (away) team improved (deteriorated) by 1.4\%. This method effectively always predicts the stronger team to win and should therefore show similar behavior as the observed outcomes. Its accuracy is always close to the limit and in one case (2012) actually exceeds it. One could explain this by the use of daily instead of end-of-season statistics but there is also another aspect in play. To describe that aspect, we need to discuss simulating seasons.

\section{Simulating seasons\label{simulating}}

With the scheduling information and the impact of chance for different pairings, we can simulate seasons in a similar manner to the Monte Carlo simulations we have discussed above, but with results that are much closer to the distribution of observed seasons. Figure \ref{2008} shows that while the simulated distribution is not equivalent to the observed one, it shows very similar trends. In addition, while the KS test does not reject any of the three simulated distributions, the distance of the one resulting from our approach to the observed one is lower than for the two Monte Carlo simulated ones.

The figure shows the result of simulating the season $10,000$ times, leading to the stabilization of the distribution. For fewer iterations, e.g. $100$ or less, distributions that diverge more from the observed season can be created. In particular, this allows the exploration of counterfactuals: if certain outcomes were due to chance, how would the model change if they came out differently? Finally, the information encoded in the different clusters -- means of statistics and co-variance matrices -- allows the generation of synthetic team instances that fit the cluster (similar to value imputation), which in combination with scheduling information could be used to generate wholly synthetic seasons to augment the training data used for learning predictive models. We plan to explore this direction in future work.

\section{Finding a good clustering\label{optimization}}

Coming back to predictive limits, there is no guarantee that the number of clusters found by the unconstrained EM will actually result in a distribution of win percentages that is necessarily close to the observed one. Instead, we can use the approach outlined in the preceding section to find a good clustering to base our chance and predictive accuracy limits on: 
\begin{enumerate}
 \item We let EM cluster teams for a fixed number of clusters (we evaluated 4--20)
 \item For a derived clustering, we simulate 10,000 seasons
 \item The resulting distribution is compared to the observed one using the Kolmogorov-Smirnov score
\end{enumerate}

The full details of the results of this optimization are too extensive to show here but what is interesting to see is that a) increasing the number of clusters does not automatically lead to a better fit with the observed distribution, and b) clusterings with different numbers of clusters occasionally lead to the same KS, validating our comment in Footnote 5.

Based on the clustering with the lowest KS, we calculate chance and predictive limit and show them in the second set of rows of Table \ref{limits}. There are several seasons for which EM already found the opimal assigment of teams to clusters (2009, 2011). Generally speaking, optimizing the fit allows to lower the KS quite a bit and leads to lower estimated chance and higher predictive limits. For both categories, however, the fact remains that different seasons were influenced by chance to differing degrees and therefore different limits exist. Furthermore, the limits we have found stay significantly below 80\% and are different from the limits than can be derived from MC simulation.

Those results obviously come with some caveats:
\begin{enumerate}
 \item Teams were described in terms of adjusted efficiencies and Four Factors -- adding or removing statistics could lead to different numbers of clusters and different cluster memberships.
 \item Predictive models that use additional information, e.g. experience of players, or networks models for drawing comparisons between teams that did not play each other, can exceed the limits reported in Table \ref{limits}.
\end{enumerate}
The table also indicates that it might be less than ideal to learn from preceding seasons to predict the current one (the approach we have chosen in our previous work): having a larger element of chance (e.g. 2009) could bias the learner against relatively stronger teams and lead it to underestimate a team's chances in a more regular season (e.g. 2010).

\section{Summary and conclusions}

In this paper, we have considered the question of the impact of chance on the outcome of (semi-)professional sports matches in more detail. In particular, we have shown that the unbiased MC simulations used to assess chance in the NFL and NHL are not applicable to the college basketball setting. We have argued that the resulting limits on predictive accuracy rest on simplifying and idealized assumptions and therefore do not help in assessing the performance of a predictive model on a particular season.

As an alternative, we propose clustering teams' statistical profiles and re-encoding a season's schedule in terms of which clusters play against each other. Using this approach, we have shown that college basketball seasons violate the assumptions of the unbiased MC simulation, given higher estimates for chance, as well as tighter limits for predictive accuracy.

There are several directions that we intend to pursue in the future. First, as we have argued above, NCAA basketball is not the only setting in which imbalanced schedules occur. We would expect similar effects in the NFL, and even in the NBA, where conference membership has an effect. What is needed to explore this question is a good statistical representation of teams, something that is easier to achieve for basketball than football/soccer teams.

In addition, as we have mentioned in Section \ref{simulating}, the exploration of counterfactuals and generation of synthetic data should help in analyzing sports better. We find a recent paper \cite{ohgraphical} particularly inspirational, in that the authors used a detailed simulation of substitution and activity patterns to explore alternative outcomes for an NBA playoff series.

Finally, since we can identify different cluster pairings and the differing of chance therein, separating those cases and training classifiers idependently for each could improve classification accuracy. To achieve this, however, we will need solve the problem of clustering statistical profiles over the entire season -- which should also allow to identify certain trends over the course of seasons.

\bibliographystyle{splncs03}
\bibliography{bibliographie.bib}

\begin{thebibliography}{1}
\providecommand{\url}[1]{\texttt{#1}}
\providecommand{\urlprefix}{URL }

\bibitem{burke07LuckNFL01}
Burke, B.: Luck and nfl outcomes 1.
  \url{http://archive.advancedfootballanalytics.com/
  2007/08/luck-and-nfl-outcomes.html}, {A}ccessed 22/06/2015

\bibitem{weka}
Frank, E., Witten, I.H.: Data Mining: Practical Machine Learning Tools and
  Techniques with Java Implementations. Morgan Kaufmann (1999)

\bibitem{ohgraphical}
Oh, M.h., Keshri, S., Iyengar, G.: Graphical model for baskeball match
  simulation. In: MIT Sloan Conference (2015)

\bibitem{kenpom}
Pomeroy, K.: Advanced analysis of college basketball. \url{http://kenpom.com}

\bibitem{weissbock13mlForNHL02}
Weissbock, J.: Theoretical predictions in machine learning for the nhl: Part
  ii.
  \url{http://nhlnumbers.com/2013/8/6/
theoretical-predictions-in-machine-learning-for-the-nhl-part-ii},
  {A}ccessed 22/06/2015

\bibitem{DBLP:conf/ai/WeissbockI14}
Weissbock, J., Inkpen, D.: Combining textual pre-game reports and statistical
  data for predicting success in the national hockey league. In: 27th Canadian
  {AI} 2014, Montr{\'{e}}al, QC, Canada, May 6-9, 2014. pp. 251--262 (2014)

\bibitem{zimmermann2013predicting}
Zimmermann, A., Moorthy, S., Shi, Z.: Predicting college basketball match
  outcomes using machine learning techniques: some results and lessons learned
  (originally in "{MLSA13}", workshop at {ECML/PKDD} 2013). arXiv preprint
  arXiv:1310.3607  (2013)

\end{thebibliography}

\appendix

\section{Clustered schedules for different seasons, unconstrained EM}

\begin{table}[th]
\centering
\begin{tabular}{c|c|c|c|c||c}
&Cluster 1 &Cluster 2 &Cluster 3 &Cluster 4 &Weaker opponent\\\hline
Cluster 1&133/197&46/182&105/272&1/45&0/696 (0.0000)\\
Cluster 2&210/227&231/352&262/374&76/247&472/1200 (0.3933)\\
Cluster 3&261/308&192/357&409/663&56/261&453/1589 (0.2851)\\
Cluster 4&210/211&341/374&424/448&515/818&975/1851 (0.5267)\\
\end{tabular}
\caption{Wins and total matches for different cluster pairings, 2009}
\end{table}

\begin{table}[th]
\centering
\begin{scriptsize}
\begin{tabular}{c|c|c|c|c|c|c||c}
&Cluster 1 &Cluster 2 &Cluster 3 &Cluster 4 &Cluster 5 &Cluster 6 &Weaker opponent\\\hline
Cluster 1&129/204&18/104&6/14&47/126&33/145&0/18&0/611 (0.0000)\\
Cluster 2&163/167&269/437&76/105&255/292&134/195&73/249&628/1445 (0.4346)\\
Cluster 3&29/34&64/95&12/18&49/58&21/41&30/87&163/333 (0.4895)\\
Cluster 4&109/136&71/240&19/46&159/232&55/119&6/87&109/860 (0.1267)\\
Cluster 5&147/163&87/166&14/23&101/123&71/118&17/57&349/650 (0.5369)\\
Cluster 6&120/120&336/361&100/117&169/172&133/141&360/579&858/1490 (0.5758)\\
\end{tabular}
\end{scriptsize}
\caption{Wins and total matches for different cluster pairings, 2010}
\end{table}

\begin{table}[th]
\centering
\begin{scriptsize}
\begin{tabular}{c|c|c|c|c|c|c|c||c}
&Cluster 1 &Cluster 2 &Cluster 3 &Cluster 4 &Cluster 5 &Cluster 6 &Cluster 7 &Weaker opponent\\\hline
Cluster 1&89/138&140/174&40/40&69/73&93/185&97/103&86/99&525/812 (0.6466)\\
Cluster 2&66/148&235/369&70/71&141/167&29/121&166/206&118/176&495/1258 (0.3935)\\
Cluster 3&2/14&15/55&29/39&16/42&0/8&20/85&4/31&0/274 (0.0000)\\
Cluster 4&10/48&48/151&36/40&42/85&2/28&55/100&28/68&91/520 (0.1750)\\
Cluster 5&166/217&187/206&43/43&79/80&205/339&80/83&148/160&703/1128 (0.6232)\\
Cluster 6&11/49&76/178&77/88&72/97&7/47&94/151&34/65&183/675 (0.2711)\\
Cluster 7&29/82&97/160&57/58&59/72&30/125&74/92&79/127&287/716 (0.4008)\\
\end{tabular}
\end{scriptsize}
\caption{Wins and total matches for different cluster pairings, 2011}
\end{table}

\begin{table}[th]
\centering
\begin{tabular}{c|c|c|c|c||c}
&Cluster 1 &Cluster 2 &Cluster 3 &Cluster 4 &Weaker opponent\\\hline
Cluster 1&108/201&110/320&20/119&19/121&0/761 (0.0000)\\
Cluster 2&362/416&610/960&105/354&175/394&362/2124 (0.1704)\\
Cluster 3&197/197&458/500&264/418&191/251&846/1366 (0.6193)\\
Cluster 4&179/191&373/454&111/245&163/258&552/1148 (0.4808)\\
\end{tabular}
\caption{Wins and total matches for different cluster pairings, 2012}
\end{table}

\begin{table}[th]
\centering
\begin{tabular}{c|c|c|c||c}
&Cluster 1 &Cluster 2 &Cluster 3 &Weaker opponent\\\hline
Cluster 1&507/807&89/374&272/567&0/1748 (0.0000)\\
Cluster 2&569/607&622/967&518/578&1087/2152 (0.5051)\\
Cluster 3&435/611&119/381&358/572&435/1564 (0.2781)\\
\end{tabular}
\caption{Wins and total matches for different cluster pairings, 2013}
\end{table}

\end{document}